# Comment: The Essential Role of Pair Matching in Cluster-Randomized Experiments, with Application to the Mexican Universal Health Insurance Evaluation

**Kai Zhang and Dylan S. Small**

## 1. INTRODUCTION

We congratulate Imai, King and Nall on a valuable paper which will help to improve the design and analysis of cluster randomized studies. The authors make two key contributions: (1) they propose a design-based estimator for matched pair cluster randomized studies that in many circumstances is a better estimator than the harmonic mean estimator; (2) they present convincing evidence that the matched pair design, when accompanied with good inference methods, is more powerful than the unmatched pair design and should be used routinely.

In this discussion, we would like to contribute our thoughts on how to construct the matched pairs. Greevy, Lu, Silber and Rosenbaum (2004) point out that in most randomized studies, only one or two variables are used in constructing the pairs. To remedy this, Greevy et al. present a method for optimal multivariate matching. They demonstrate in an example with 14 covariates and 132 units that the optimal matching achieves substantially better balance on all 14 covariates than an unmatched design.

*Kai Zhang is Ph.D. student, Department of Statistics, The Wharton School, University of Pennsylvania, 400 Huntsman Hall, 3730 Walnut St., Philadelphia, Pennsylvania 19104, USA e-mail: zhangk@wharton.upenn.edu. Dylan S. Small is Associate Professor, Department of Statistics, The Wharton School, University of Pennsylvania, 400 Huntsman Hall, 3730 Walnut St., Philadelphia, Pennsylvania 19104, USA e-mail: dsmall@wharton.upenn.edu.*

Greevy et al. considered the situation in which we want to use all available units in the experiment. In cluster randomized studies, because of cost considerations, we can often only use some of the clusters, that is, there are $N = 2k$ clusters but we would only like to include $2m$ ($m < k$) clusters in the experiment; see Murray (1998), Chapter 10, for several examples. How should we choose the best $m$ matched pairs? In our discussion, we compare several methods of constructing matched pairs for this setting.

Our discussion is organized in the following way. We introduce and discuss four methods of matching in the next section. Then we conduct simulations to compare these methods and the results are summarized in Section 3 and 4. Conclusions of our findings are given in Section 5.

## 2. FOUR METHODS OF MATCHING PAIRS

The goal of matching is to produce a design for the experiment that has high power relative to other designs. Matching methods seek to do this by defining a distance between every pair of units and then making the total distance between the matched units as small as possible. One distance for matching is the Mahalanobis distance (Rubin, 1979). We will compare matching methods by comparing the total Mahalanobis distance between the matched units.

We consider four methods of constructing $m$ matched pairs when there are $N = 2k$ ($k > m$) units. Three of the methods make use of the optimal nonbipartite matching algorithm described by Greevy et al. (2004) which, for a given $2r$ units and a $(2r) \times (2r)$ distance matrix, returns the $r$ pairs which minimize the total distance between the units in the pairs.

1. *The random method*. A simple random sample of $2m$ units from the $N$ units is selected. The

selected units are matched optimally using the optimal nonbipartite matching algorithm.

2. *The ranking method*. Optimal nonbipartite matching is applied to construct the $k$ pairs from all $2k$ units which minimize the total distance between matched pairs. Then, the $m$ pairs which have the smallest distance are selected. King et al. (2007) use a variant of the ranking method in choosing which clusters to conduct individual-level surveys in.
3. *The greedy method*. First, pick the pair with the smallest distance over all pairs and then remove the two units in this pair from consideration. Then pick the pair with the smallest distance over all remaining possible pairs, and remove the two units in this pair from consideration. Continue until $m$ pairs have been selected.
4. *The optimal method*. The optimal method minimizes the total distance within the $m$ chosen pairs. The procedure is as follows: (1) create $N - 2m$ *artificial sinks*: the Mahalanobis distance between any two artificial sinks is set to be large (ideally infinity) while the distance between any artificial sink and any real unit is set to be zero; (2) find the optimal matching of $N - m$ pairs of the combined $2N - 2m$ units (the real units plus the artificial sinks) using optimal nonbipartite matching; and (3) select the $m$ pairs that consist of two real units. The reasons that this method chooses the optimal $m$ pairs are the following. Since the distance between a real unit and an artificial sink is 0 while the distance between any two artificial sinks is very large, each artificial sink will be matched to a real unit. Therefore, $N - 2m$ pairs of real units and artificial sinks will be constructed and the remaining $2m$ real units will form pairs that we select. Since the distances between the artificial sinks and the real units are 0, the total distance of the $N - m$ pairs is the same as the total distance of the $m$ pairs formed by real units. Therefore, minimizing the total distance of the $N - m$ pairs is the same as minimizing the total distance of the selected $m$ pairs of real units.

The random method is the easiest thing to do. However, due to the fact that this method choose units blindly from the pool, undesirable pairs can be matched and the performance can be bad. The ranking method and the greedy method both attempt to provide the best pairs in the overall pool of units, but do not typically choose the optimal pairs.

## 3. COMPARISON OF THE FOUR METHODS THROUGH SIMULATION

Our first simulation is conducted in the following way. Let $N = 100$, $k = 50$ and $m = 10$. The data are generated with eight covariates: C1 is from an exponential distribution with mean 1; C2 is from a $t$ distribution with 3 degrees of freedom; C3 is from a normal distribution with mean 1 and variance 1; C4 is from a uniform distribution over [0, 2]; C5, C6, C7 and C8 are from a multivariate normal distribution with mean vector $(1, 1+\frac{1}{99}, 1+\frac{2}{99}, \ldots, 1+\frac{98}{99}, 2)$, and a covariance matrix that has 1's on the diagonal and 0.5 off-diagonal. C1, C2, C3, C4 are independent of each other and are independent from the last four covariates.

We ran the simulation 10,000 times and calculate the ratios of total distance between the optimal method and the three other methods. The ratios are denoted $R1$ (the random method compared to the optimal method), $R2$ (the ranking method compared to the optimal method) and $R3$ (the greedy method compared to the optimal method).

The summary of $R1$, $R2$ and $R3$ is given in Table 1 and the histograms of $R1$, $R2$ and $R3$ are given in Figure 1.

The optimal method is much better than the random method in terms of total distance; on average, the optimal method reduces the distance 50%. In all simulations, the optimal method provided at least 30% improvement and in some situations, it provided more than 70% improvement.

Compared to the ranking method (matching 50 pairs and selecting the best 10 out of 50), the optimal method gained a median of 2.5% and as much as 35%.

The greedy algorithm performed almost as well as the optimal method for this setting. However, when we want to match a higher proportion of the available units, the greedy algorithm does not perform as well; we shall discuss this later.

We conducted another simulation with a different setup where the data are skewed and have heavier tails. The eight covariates are generated as follows: C1 is from an exponential distribution with mean 1; C2 is from a $t$ distribution with 3 degrees of freedom; C3 is from a Cauchy distribution; C4 is from a uniform distribution over [0, 2]; C5, C6, C7 and C8 are from exponentiating the multivariate normal data generated as in the previous setup. C1, C2, C3, C4 are independent of each other and are independent from the last four covariates.

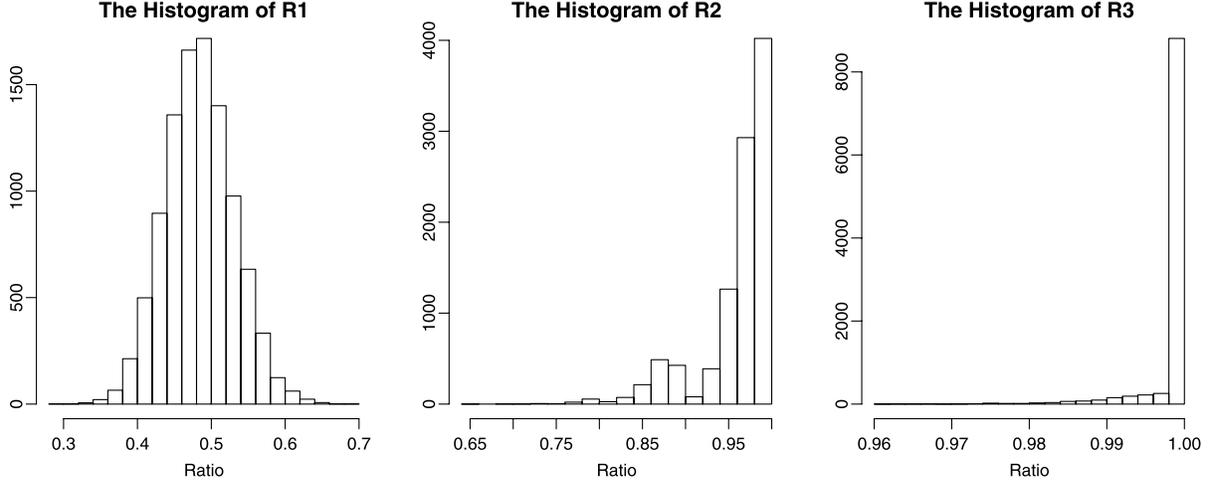

Fig. 1. *Histograms of $R1$, $R2$ and $R3$ in the first case.*

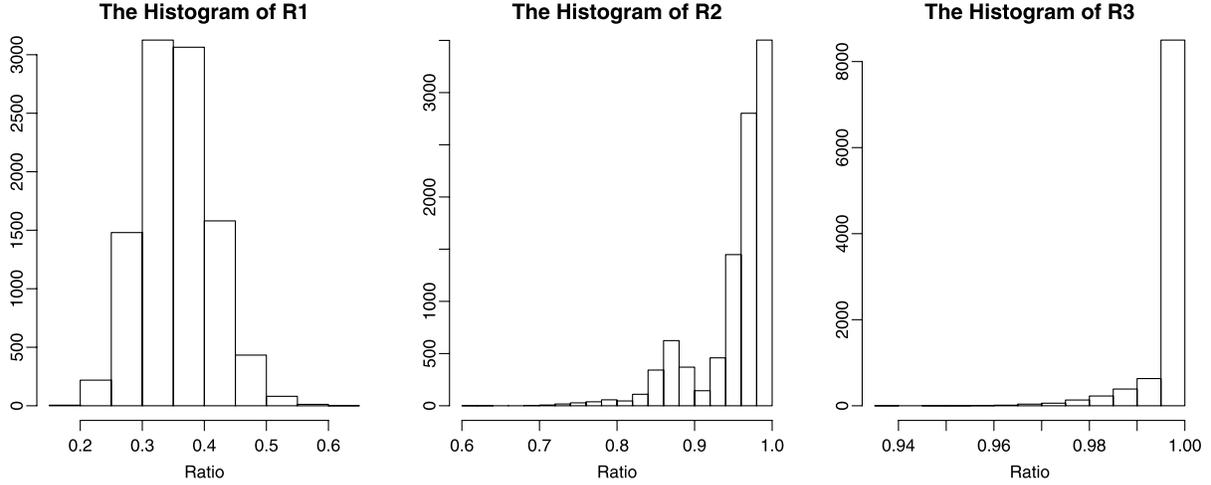

Fig. 2. *Histograms of $R1$, $R2$ and $R3$ over three kinds of data.*

The summary of $R1$, $R2$ and $R3$ is given in Table 2 and histograms of $R1$, $R2$ and $R3$ are given in Figure 2.

We can see that compared to the previous simulation, we gain more in all three ratios. The gains in $R1$ and $R2$ are more significant.

Our simulation shows that the optimal method is far better than the random method. Also, the optimal method is always at least as good as the ranking method and can be much better in some situations. Finally, the optimal method is always at least as good as the greedy method but they have similar performance.

## 4. COMPARISON OF THE OPTIMAL METHOD AND THE GREEDY METHOD

We know that the optimal method gives the best $m$ pairs among $N = 2k$ clusters in all circumstances and for all $m$. However, the greedy method performs almost as well as the optimal method in the situations discussed above, namely the gain from utilizing the optimal method is only about 1%. In this section, we investigate situations when the greedy method does not perform as well. Again in this section we simulate 10,000 times for each case.

We first consider matching methods when $m = 30$, $m = 45$ and $m = 50$, where $N$ always equals 100, over three kinds of single-covariate data here. (1) One covariate from a standard Cauchy distribution. (2) One covariate from a standard normal distribution. (3) One covariate from a uniform distribution over $[0,1]$. The results are shown in Table 3 and Figure 3.

We see that when $m = 30$, the optimal method gains most from the Cauchy data, but only 3.1% on average. The optimal method gains 1.2% and 0.5% on average when $m = 30$ from the normal data and the uniform data, respectively. When $m = 45$, the optimal method gains most from the Cauchy data too, with 17.0% on average. The optimal method gains 11.1% and 8.7% on average when $m = 45$ from the normal data and the uniform data, respectively.

When $m = 50$, we see a different pattern in which the gain is greater for the normal and uniform data than the Cauchy. For the Cauchy data, the average gain is 16.4%, while for the normal data and the uniform data, the gains are 42.1% and 44.2%, respectively.

We also investigate cases when there are two covariates in the data. We take $N = 100$ and $m = 50$ and consider twelve cases. In the first three cases we consider (1) independent covariates generated from the standard Cauchy distribution, (2) independent covariates generated from the standard normal distribution and (3) independent covariates generated from the uniform distribution over $[0, 1]$. We consider nine other cases in which the covariates are independently generated from discrete multimodal distributions. The probability mass functions are summarized in the table below.

The summary of ratios of total distances is given in the boxplots in Figure 4 and the histograms in Figure 5.

We observe that the gains of the optimal method when the underlying distributions are normal or uniform are around 20%, which is not as much as the 40% gain in the one covariate case. We also observe that as the number of modes increases, the gains become greater and the histograms of ratios become more similar to the ones of the normal and uniform distributions (Cases 2 and 3).

## 5. CONCLUSION

Imai et al. have made an important contribution to the design and analysis of cluster randomized trials by showing the advantages of a matched pair design compared to an unmatched pair design and providing an inference method that is appropriate for a matched pair design. In our discussion, we have considered how to construct the matched pairs when we only want to use a proportion of the pool of available clusters in the experiment. We have shown by simulation that it is very important to consider all available clusters when constructing the matched pairs rather than randomly choosing some to focus on. Among the methods which focus on all available clusters, the ranking method and the greedy method perform acceptably but the optimal method can be substantially better than them in some situations. Consequently, we recommend use of the optimal method in constructing matched pairs.

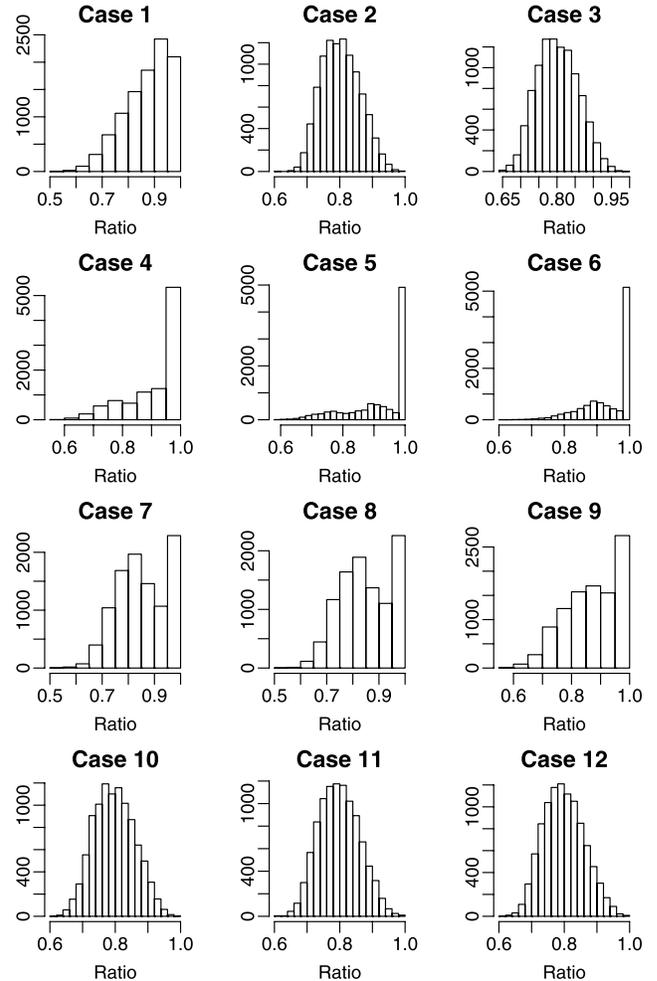

FIG. 5. *Histograms of ratios for cases in Table 4.*

## ACKNOWLEDGMENTS

We are grateful for Dongyu Lin's and Paul Rosenbaum's insightful suggestions.

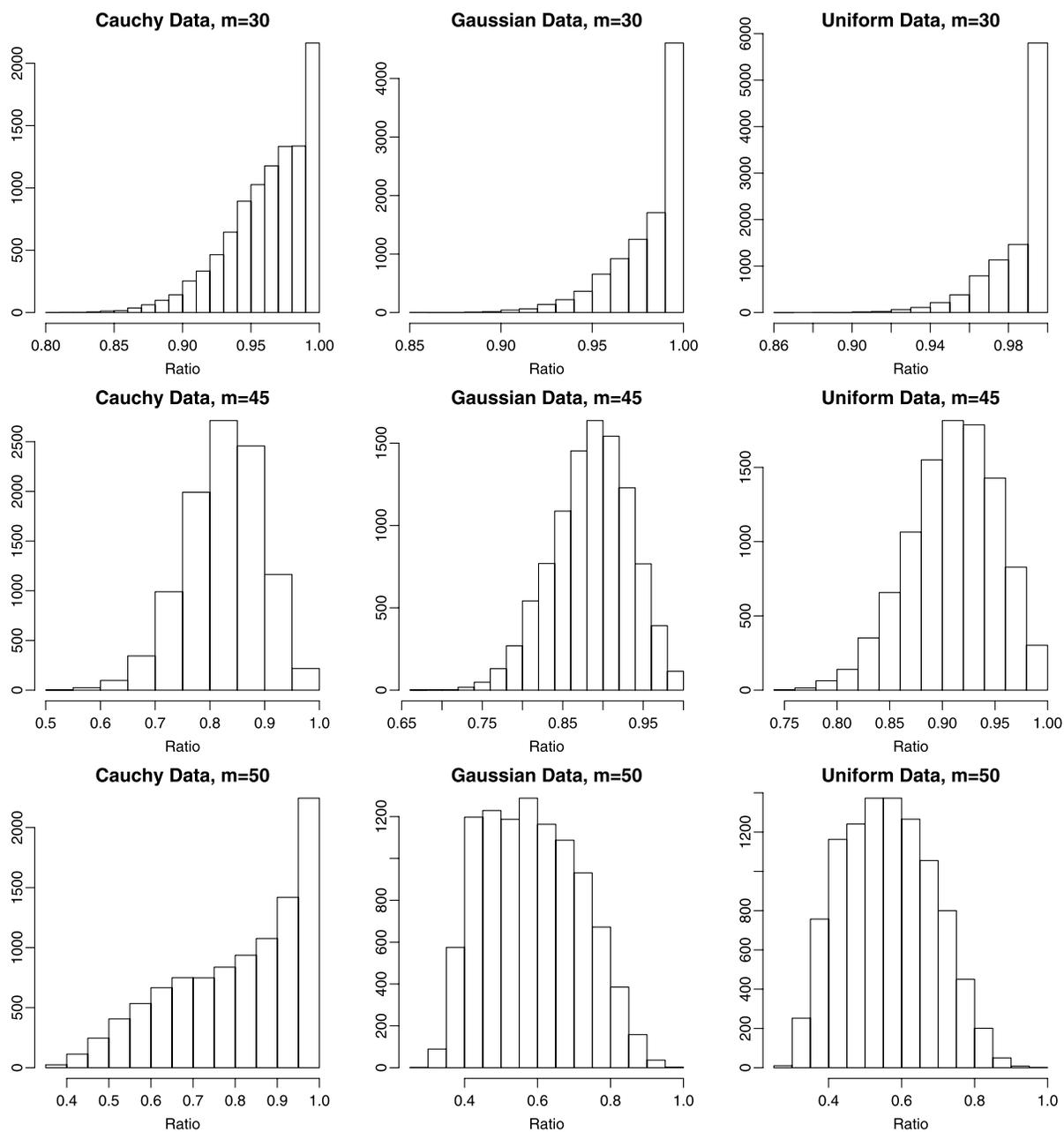

Fig. 3. *Histograms of ratios over three kinds of data.*

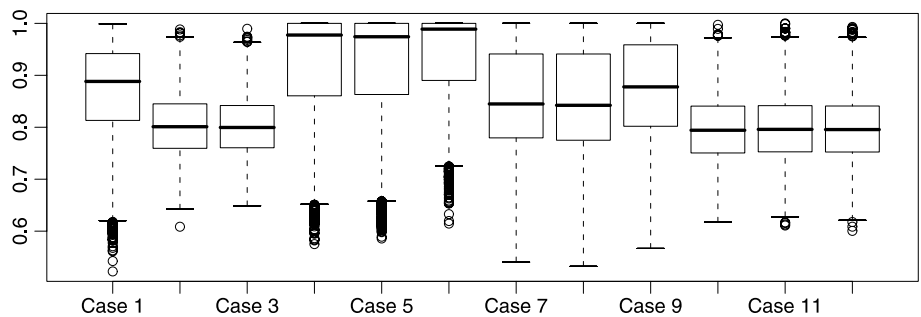

Fig. 4. *Boxplots of ratios for cases in Table 4.*

TABLE 1
*Summary of R1, R2 and R3 in the first case*

|    | Minimum | 25% quantile | Median | Mean | 75% quantile | Maximum |
|----|---------|--------------|--------|------|--------------|---------|
| $R1$ | 0.2927 | 0.4523 | 0.4832 | 0.4842 | 0.5148 | 0.6818 |
| $R2$ | 0.6517 | 0.9535 | 0.9744 | 0.9603 | 0.9879 | 1.0000 |
| $R3$ | 0.9604 | 1.0000 | 1.0000 | 0.9989 | 1.0000 | 1.0000 |

TABLE 2
*Summary of R1, R2 and R3 in the second case*

|    | Minimum | 25% quantile | Median | Mean | 75% quantile | Maximum |
|----|---------|--------------|--------|------|--------------|---------|
| $R1$ | 0.1763 | 0.3152 | 0.3524 | 0.3548 | 0.3923 | 0.6029 |
| $R2$ | 0.6146 | 0.9455 | 0.9703 | 0.9536 | 0.9862 | 1.0000 |
| $R3$ | 0.9373 | 1.0000 | 1.0000 | 0.9978 | 1.0000 | 1.0000 |

TABLE 3
*Summary of ratios over three kinds of data*

|    | $m$ | Minimum | 25% quantile | Median | Mean | 75% quantile | Maximum |
|----|-----|---------|--------------|--------|------|--------------|---------|
| Cauchy data   | 30 | 0.8087 | 0.9451 | 0.9687 | 0.9631 | 0.9873 | 1.0000 |
| Gaussian data | 30 | 0.8510 | 0.9706 | 0.9879 | 0.9816 | 1.0000 | 1.0000 |
| Uniform data  | 30 | 0.8696 | 0.9782 | 0.9945 | 0.9869 | 1.0000 | 1.0000 |
| Cauchy data   | 45 | 0.5484 | 0.7787 | 0.8291 | 0.8244 | 0.8747 | 0.9919 |
| Gaussian data | 45 | 0.6777 | 0.8543 | 0.8884 | 0.8853 | 0.9200 | 1.0000 |
| Uniform data  | 45 | 0.7434 | 0.8831 | 0.9128 | 0.9103 | 0.9406 | 1.0000 |
| Cauchy data   | 50 | 0.3503 | 0.6847 | 0.8357 | 0.8036 | 0.9424 | 1.0000 |
| Gaussian data | 50 | 0.2966 | 0.4755 | 0.5786 | 0.5848 | 0.6849 | 0.9854 |
| Uniform data  | 50 | 0.2747 | 0.4637 | 0.5576 | 0.5612 | 0.6524 | 0.9611 |

TABLE 4
*Probability mass functions in considered cases*

|    | Possible values | Probability | Number of modes |
|----|-----------------|-------------|-----------------|
| Case 4  | $(-10, -1, 1, 10)$ | $(0.4, 0.1, 0.1, 0.4)$ | 2 |
| Case 5  | $(-10, -1, 1, 10)$ | $(0.1, 0.4, 0.1, 0.4)$ | 2 |
| Case 6  | $(-10, -1, 2, 20)$ | $(0.6, 0.1, 0.2, 0.1)$ | 2 |
| Case 7  | $(-20, -10, -1, 1, 10, 20)$ | $(\frac{1}{12}, \frac{1}{3}, \frac{1}{12}, \frac{1}{12}, \frac{1}{3}, \frac{1}{12})$ | 2 |
| Case 8  | $(-20, -10, -1, 1, 10, 20)$ | $(\frac{1}{12}, \frac{1}{3}, \frac{1}{12}, \frac{1}{12}, \frac{1}{12}, \frac{1}{3})$ | 2 |
| Case 9  | $(-20, -10, -1, 2, 20, 40)$ | $(\frac{1}{12}, \frac{1}{3}, \frac{1}{12}, \frac{1}{12}, \frac{1}{12}, \frac{1}{3})$ | 2 |
| Case 10 | $(-10, -10 + \frac{20}{19}, \ldots, 10)$ | $(\frac{1}{50}, \frac{4}{50}, \frac{4}{50}, \frac{1}{50}, \frac{1}{50}, \frac{4}{50}, \ldots, \frac{1}{50})$ | 10 |
| Case 11 | $(-10, -10 + \frac{20}{19}, \ldots, 10)$ | $(\frac{1}{50}, \frac{4}{50}, \frac{1}{50}, \frac{4}{50}, \ldots, \frac{4}{50})$ | 10 |
| Case 12 | $(-10, -10 + \frac{40}{19}, \ldots, 30)$ | $(\frac{4}{50}, \frac{1}{50}, \frac{4}{50}, \frac{1}{50}, \ldots, \frac{1}{50})$ | 10 |